\documentclass[12pt]{article}
\def\be{\begin{equation}}
\def\ee{\end{equation}}
\def\bea{\begin{eqnarray}}
\def\eea{\end{eqnarray}}
\usepackage{graphicx}

\catcode`\@=11
\def\lsim{\mathrel{\mathpalette\@versim<}}
\def\gsim{\mathrel{\mathpalette\@versim>}}
\def\@versim#1#2{\vcenter{\offinterlineskip
\ialign{$\m@th#1\hfil##\hfil$\crcr#2\crcr\sim\crcr } }}
\catcode`\@=12
\usepackage{axodraw}

\catcode`@=12
\begin{document}

\thispagestyle{empty}
\begin{flushright}
UCRHEP-T518\\
\today\
\end{flushright}
\vspace{0.3in}
\begin{center}
{\LARGE \bf Dirac neutrino mass generation from dark
matter\\}
\vspace{1.0in}
{\bf Yasaman Farzan$^1$ and Ernest Ma$^{2,3}$\\}
\vspace{0.2in}
{\sl $^1$ School of Physics, Institute for Research in Fundamental Sciences
(IPM),\\
P.~O.~Box 19395-5531, Tehran, Iran\\}
\vspace{0.1in}
{\sl $^2$ Department of Physics and Astronomy, University of California,\\
Riverside, California 92521, USA\\}
\vspace{0.1in}
{\sl $^3$ Kavli Institute for the Physics and Mathematics of the Universe
(IPMU),\\ University of Tokyo, Kashiwa 277-8583, Japan\\}
\end{center}
\vspace{1.0in}
\begin{abstract}\
In 2006,    a simple extension of the Standard Model was proposed
in which neutrinos obtain  radiative Majorana masses at one-loop
level  from their couplings with dark matter, hence the term
``scotogenic,'' from the Greek ``scotos'' meaning darkness.  Here
an analogous mechanism for Dirac neutrino masses is discussed in a
minimal model. In different ranges of the parameter space, various
candidates for dark matter are possible. In particular, the
lightest  Dirac fermion which appears in the loop diagram
generating  neutrino mass can be  a viable dark matter candidate.
Such a possibility  does not exist for the Majorana case.
Realistic neutrino mixing in the context of $A_4$ is discussed.  A
possible supersymmetric extension is also briefly discussed.
\end{abstract}

\newpage
\baselineskip 24pt

Dirac neutrino masses have not received much attention in the
literature mainly  because of  two reasons: (1) In the Standard
Model (SM) of particle interactions, there are left-handed lepton
doublets $(\nu,l)_L$ and right-handed charged-lepton singlets
$l_R$ but no $\nu_R$ because it transforms trivially under the
$SU(3)_C \times SU(2)_L \times U(1)_Y$ gauge symmetry and there is
no need for its existence.  If it is added in by hand,  the
neutrino can then obtain a Dirac mass $m_D$ in the same way as all
the other fermions (quarks and charged leptons), i.e. from the
vacuum expectation value of the  scalar Higgs doublet of the
Standard Model. However, since $\nu_R$ is a neutral singlet, there
is no symmetry which prevents it from having a large Majorana mass
$M$. As a result, $\nu_L$ obtains an effective small Majorana mass
from the seesaw mechanism~\cite{seesaw}, i.e. $m_\nu \simeq
-m_D^2/M$. (2) If a symmetry is imposed in such a way that the
lepton number is conserved,  the Majorana mass term for $\nu_R$
will be forbidden. In that case, because neutrino masses are known
to be of order 1 eV or less, the corresponding Yukawa couplings
must be of order $10^{-11}$ or smaller.  Such a small value is
considered by many to be intrinsically unacceptable.

Nevertheless, up to now, there is  not any indisputable evidence
for  the Majorana nature of the neutrinos from the searches for
the neutrinoless double beta decay. Thus, the possibility of Dirac
neutrino masses cannot be discounted.  To overcome the above
theoretical objections, it is proposed in this paper that
neutrinos are Dirac fermions, with two important properties. (1)
They are protected from becoming Majorana fermions by a
$U(1)_{B-L}$ global or gauge symmetry. (2) They are protected from
having a tree-level mass by a $Z_2$ symmetry which is identifiable
with that of dark matter, as well as another $Z_2$ symmetry which
sets them apart from other Dirac fermions.  The latter symmetry is
broken explicitly by soft terms.  It may also be replaced by
supersymmetry, but that would require a much larger Higgs content.
As a result, neutrinos acquire one-loop radiative masses through
their couplings with dark matter, hence the term ``scotogenic,''
from the Greek ``scotos'' meaning darkness. These Dirac neutrino
masses can be highly suppressed in the same way that the usual
seesaw Majorana neutrino masses are highly suppressed. Their
smallness can be also explained by the smallness of the soft terms
breaking the $Z_2$ symmetry.

In 2006, it was proposed~\cite{m06,Celine} that neutrinos are
massive only because of their couplings with dark matter.  This
idea connects two of the most important issues in the particle
physics and astrophysics.  The idea was easily
implemented~\cite{m06} in a simple extension of the Standard Model
by adding a second scalar doublet $(\eta^+,\eta^0)$ and three
neutral singlet fermions $N_i$ which are odd under an extra
exactly conserved discrete $Z_2$ symmetry~\cite{dm78}, in analogy
to the $R$ parity in supersymmetry.  As a result, either $\eta_R =
\sqrt{2} Re(\eta^0)$ or the lightest $N$ may be considered a
candidate for dark matter. In particular $\eta$ has been called
the ``inert'' scalar doublet in a model proposed~\cite{bhr06}
after Ref.~\cite{m06} and studied by many authors
since then~\cite{lhy11}. Variations of the original idea also abound
and have become an active area of
research~\cite{gs08-1,f09,ms09,m09,p11,st11,kns12}.

In almost all previous such applications, neutrino masses have
always been assumed to be of Majorana type.  Suppose they are
exactly Dirac. Is the connection between neutrino mass and dark
matter still possible? If so, what are the necessary theoretical
ingredients for it to happen, and what are the phenomenological
consequences?  In~\cite{kns11}, using scalar singlets, a radiative
Dirac neutrino mass  is obtained; however, in this mechanism, the
dark matter fields do not propagate in the loop.
 Employing an idea similar to that proposed in
Ref.~\cite{m06}, Ref~\cite{gs08} suggests a model both for a dark
matter candidate and generation of radiative Dirac neutrino  mass.
As indicated below, this model shares some features with the model
introduced in the present paper. In \cite{kns12}, a model is
introduced in which neutrinos obtain a Dirac mass via a one-loop
diagram similar to that in \cite{gs08} and a Majorana mass via
two-loop diagrams after spontaneous breaking of the lepton number
symmetry. In our model described below, the neutrino mass is
purely of the Dirac type.

Consider first the imposition of a conserved additive lepton
number to protect the neutrino mass from becoming Majorana.  We
choose to do so by extending the Standard Model to include $B-L$
as either a global or gauged $U(1)$ symmetry.  The latter has long
been known to be a well-motivated anomaly-free extension which
requires the existence of three singlet right-handed neutrinos. Of
course, in breaking the gauged $U(1)_{B-L}$, we have to be sure
that the global $U(1)_{B-L}$ symmetry of the sector
relevant to the present study remains intact. This can be done
easily by a scalar field transforming under $U(1)_{B-L}$ but not
coupling to other fields with nonzero $B-L$. The second step is to
forbid a tree-level Dirac neutrino mass $m_\nu$, and yet allow a
tree-level charged-lepton mass $m_l$. To do this, the simplest way
is to impose a $Z_2^{(A)}$ symmetry such that $\nu^c$ is odd but
all other fermions are even. There is therefore no connection
between $\nu$ and $\nu^c$ at the tree level.  To make them connect
in one loop, new particles are postulated which are odd under an
exactly conserved $Z_2^{(B)}$, and the previous $Z_2^{(A)}$ is
allowed to be broken by soft terms.  Another way is to make the
model supersymmetric as well so that $m_l$ comes from $\Phi_1 =
(\phi_1^0,\phi_1^-)$ but $m_\nu$ is forbidden to couple to $\Phi_2
= (\phi_2^+,\phi_2^0)$ which is assumed odd under $Z_2^{(B)}$.  In
either case, we need to add  heavy neutral singlet Dirac fermions
$(N_i,N^c_i)$ of odd $Z_2^{(B)}$ transforming under $U(1)_{B-L}$
and a neutral singlet scalar $\chi^0$ of odd $Z_2^{(B)}$ which is
trivial under $U(1)_{B-L}$.

First, let us consider the minimal non-supersymmetric model.  It
is a simple extension of the Standard Model in the same spirit of
Ref.~\cite{m06}.  Its particle content is listed in Table 1.   In
addition to the usual particles of the Standard Model, we have
added three copies of the Weyl spinors $\nu^c$, three copies of
the Dirac spinor pairs $(N,N^c)$, one extra scalar doublet
$\eta=(\eta^+, \eta^0)$ and one real scalar $\chi^0$.  The $B-L$
symmetry prevents $N$, $N^c$ as well as $\nu^c$ from having a
Majorana mass.
\begin{table}[t]
\begin{center}
\begin{tabular}{|c|c|c|c|c|c|c|}
\hline
particles & $SU(3)_C$ & $SU(2)_L$ & $U(1)_Y$ & $U(1)_{B-L}$ &
$Z_2^{(A)}$ & $Z_2^{(B)}$ \\
\hline
$(u,d)$ & 3 & 2 & 1/6 & 1/3 & + & +  \\
$u^c$ & $3^*$ & 1 & --2/3 & --1/3 & + & + \\
$d^c$ & $3^*$ & 1 & 1/3 & --1/3 & + & + \\
\hline
$(\nu,e)$ & 1 & 2 & --1/2 & --1 & + & + \\
$e^c$ & 1 & 1 & 1 & 1 & + & + \\
$\nu^c$ & 1 & 1 & 0 & 1 & -- & + \\
\hline
$(\phi^+,\phi^0)$ & 1 & 2 & 1/2 & 0 & + & + \\
$(\eta^+,\eta^0)$ & 1 & 2 & 1/2 & 0 & + & -- \\
$\chi^0$ & 1 & 1 & 0 & 0 & -- & -- \\
\hline
$N$ & 1 & 1 & 0 & --1 & + & -- \\
$N^c$ & 1 & 1 & 0 & 1 & + & -- \\
\hline
\end{tabular}
\caption{Assignments of the particles of the
minimal model under $B-L$, $Z_2^{(A)}$ and $Z_2^{(B)}$.}
\end{center}
\end{table}
Note that $Z_2^{(A)}$ is broken softly by the trilinear term $A
\chi \Phi^\dagger \eta$, whereas $Z_2^{(B)}$ remains unbroken.
($\Phi=(\phi^+ \ , \phi^0)$ is the standard model Higgs doublet.)
The one-loop Dirac neutrino mass is thus generated, as shown in
Fig.~1.

Note that $\chi^0$ is essential here for a scotogenic Dirac
neutrino mass, whereas the scalar singlet considered in
Ref.~\cite{f09} is not needed for a scotogenic Majorana neutrino
mass.

\thispagestyle{empty}

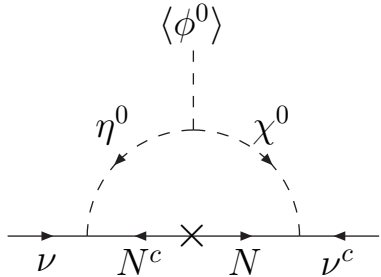
\begin{figure}[htb]
\begin{center}
\begin{picture}(260,100)(0,0)
\ArrowLine(60,10)(90,10) \ArrowLine(130,10)(90,10)
\ArrowLine(130,10)(170,10) \ArrowLine(200,10)(170,10)
\DashArrowArc(130,10)(40,90,180)5 \DashArrowArcn(130,10)(40,90,0)5
\DashLine(130,50)(130,80)5 \Text(75,0)[]{\large $\nu$}
\Text(185,0)[]{\large $\nu^c$} \Text(110,0)[]{\large $N^c$}
\Text(150,0)[]{\large $N$} \Text(100,52)[]{\large $\eta^0$}
\Text(160,52)[]{\large $\chi^0$} \Text(130,90)[]{\large $\langle
\phi^0 \rangle$} \Text(130,10)[]{\Large $\times$}
\end{picture}
\end{center}
\caption{One-loop generation of Dirac neutrino mass in the minimal
model.}
\end{figure}
Whereas a scalar singlet was discussed as dark matter by itself
many years ago~\cite{sz85,m94,bptv01}, our proposal may be
considered a natural justification of its existence.

Let the Yukawa interactions be given by $f_{\alpha k} \nu_\alpha
N^c_k \eta^0$ and $h_{k \beta} N_k \nu^c_\beta \chi^0$. Without
loss of generality, the $A$ parameter of the trilinear $A \chi
\bar{\phi}^0 \eta^0$ term may always be chosen real, as well as
the vacuum expectation value $\langle \phi^0 \rangle = v$. Let
$\eta^0 = (\eta_{R} + i\eta_{I})/\sqrt{2}$, then there is a mixing
between $\eta_{R}$ and $\chi^0$, but not between $\eta_{I}$ and
$\chi^0$.   Assuming in addition that $\eta_I$ is a mass
eigenstate and denoting the mass eigenstates of the
$(\chi^0,\eta_{R})$ sector as $\zeta_{1,2}$ with mixing angle
$\theta$, the one-loop Dirac neutrino mass matrix is then given by
\begin{equation} \label{simpleMaSS}
({\cal M}_\nu)_{\alpha \beta} = {\sin \theta \cos \theta \over 16 \pi^2
\sqrt{2}} \sum_k f_{\alpha k} h_{k \beta} m_{N_k}
\left[ {m^2_{\zeta_1} \over m^2_{\zeta_1} -
m^2_{N_k}} \ln {m^2_{\zeta_1} \over m^2_{N_k}} -
{m^2_{\zeta_2} \over m^2_{\zeta_2} -
m^2_{N_k}} \ln {m^2_{\zeta_2} \over m^2_{N_k}} \right].
\end{equation}
This is in complete analogy to that of the radiative Majorana
seesaw~\cite{m06}, with suppression of the neutrino mass from the
usual assumption of very large $m_N$ (now Dirac) as well as the
loop factor.  In addition, this diagram is only nonzero because of
the soft breaking of $Z_2^{(A)}$.  Thus, it is natural for the
parameter $A$ to be small.  In the limit $A=0$, the mixing angle
$\theta$ in the above equation would be zero.

We assume that there are three copies of $(N,N^c)$ so that all
three neutrinos obtain scotogenic Dirac masses.  If there is only
one copy, then two neutrinos will be massless, which is clearly
unrealistic. If there are two copies, one will be massless, which
is acceptable as far as present neutrino phenomenology is
concerned.  From Table~1, it can be  easily confirmed that with
three copies of $\nu^c$,  $U(1)_{B-L}$ will be anomaly-free.

In this model, $\Phi$ is the SM Higgs doublet with the usual Higgs
boson $H$ as its only physical degree of freedom.  It has the
usual SM decay modes, except for corrections due to its
interactions with $\eta$ and $\chi^0$.  For example, $H$ may decay
into $\zeta_1 \zeta_1$ if kinematically allowed.  If $\zeta_1$ is
dark matter,  this decay would then be invisible.  It would affect
the search for the SM Higgs boson, as studied already in
Ref.~\cite{cmr07}. Another possible effect is that the coupling of
$H$ to $\eta^+ \eta^-$ would change the one-loop decay of $H$ to
$\gamma \gamma$, thus affecting also the search for the SM Higgs
boson via this channel.  A third effect is the existence of the
quartic $\chi \chi \Phi^\dagger \Phi$ coupling, which may
contribute significantly to the effective potential of $H$ and
modify its stability condition as a function of mass.  It may also
induce a one-loop contribution to the $H^3 T$ term at finite
temperature to cause a first-order phase transition needed for the
electroweak baryogensis.

The couplings $f_{\alpha k} L_\alpha N_k^c \eta$ contribute to
radiative lepton flavor violating rare decays:
\begin{equation}
\Gamma(l_\alpha^- \to l_\beta^- \gamma)=\frac{m_\alpha^3}{16 \pi}
\sigma_R^2,
\end{equation}
where \begin{equation}\sigma_R= \sum_k e f_{\alpha k} f_{\beta
k}^* m_\alpha \frac{i}{16 \pi^2 m_{\eta^+}^2} \left[ \frac{t
\ln t}{2(t-1)^4}+\frac{t^2-5t
-2}{12(t-1)^3}\right]\label{sigmaR},
\end{equation}
with $t=(m_{N_k}^2/m_{\eta^+}^2)$.  For $t \to 0$, $t \to \infty$
and $t \to 1$, the combination in the last parenthesis of Eq.~(\ref{sigmaR})
converges respectively to $1/6$, $1/(12t)$ and $1/24$.  For
$m_{N_k}\gg m_{\eta^+}$, which is the seesaw limit, we find
\begin{equation}
\left(\sum_k \frac{f_{\alpha k}f_{\beta k}^*}{m_{N_k}^2}
\right)^{1/2}\sim 8\times 10^{-5} \left( \frac{{\rm B}(l_\alpha
\to l_\beta \gamma)}{10^{-12}}\right)^{1/4}{\rm GeV}^{-1}.
\end{equation}
We will consider first this scenario, so that the dark-matter candidate
of our model is the lightest of the three exotic neutral scalars:
$\zeta_{1,2}$ or $\eta_{I}$.

The most general scalar potential consisting of $\Phi$, $\eta$, and $\chi$
is given by
\begin{eqnarray}
V &=& \mu_1^2 \Phi^\dagger \Phi + \mu_2^2 \eta^\dagger \eta + {1 \over 2}
\mu_3^2 \chi^2 + {1 \over 2} \lambda_1 (\Phi^\dagger \Phi)^2 +
{1 \over 2} \lambda_2 (\eta^\dagger \eta)^2 \nonumber \\
&+& \lambda_3 (\Phi^\dagger \Phi)(\eta^\dagger \eta) + \lambda_4
(\Phi^\dagger \eta)(\eta^\dagger \Phi) + {1 \over 2} \lambda_5
(\Phi^\dagger \eta)^2 + H.c. \nonumber \\ &+& {1 \over 4} \lambda_6 \chi^4
+ {1 \over 2} \lambda_7 (\Phi^\dagger \Phi) \chi^2 + {1 \over 2}
\lambda_8 (\eta^\dagger \eta) \chi^2 + A \chi \Phi^\dagger \eta + H.c. \label{Vmain}
\end{eqnarray}
This potential preserves $Z_2^{(B)}$ and breaks $Z_2^{(A)}$ softly by
the last term.  The parameter $A$ may be chosen real by a phase rotation
of $\eta$ relative to $\Phi$, but then $\lambda_5$ is in general complex.
For simplicity, we choose it to be real so that $\eta_I$ is a mass
eigenstate and decouples from the $(\chi^0, \eta_R)$ sector.
The resulting mass spectrum is given by
\begin{eqnarray}
m^2_H &=& 2 \lambda_1 v^2, \\
m^2_{\eta^+} &=& \mu_2^2 + \lambda_3 v^2, \\
m^2_{\eta_I} &=& \mu_2^2 + (\lambda_3 + \lambda_4 - \lambda_5) v^2, \\
m^2_{(\chi,\eta_R)} &=& \pmatrix{ \mu_3^2 + \lambda_7 v^2  & \sqrt{2} A v
\cr \sqrt{2} A v & \mu_2^2 + (\lambda_3 + \lambda_4 + \lambda_5) v^2}.
\end{eqnarray}
This is very similar to previous studies such as Ref.~\cite{f09}
with an important conceptual difference.  Since the parameter $A$
breaks $Z_2^{(A)}$, it may be argued that it is small.  This
suppresses the radiative neutrino Dirac mass as well as the mixing
between $\eta_R$ and $\chi$.  Hence the dark-matter candidate of
this model can be dominantly a singlet and as a result, it can
naturally evade the constraints from the electroweak interactions
of a doublet. If there is no $Z_2^{(A)}$ symmetry,  the mixing
between $\eta_R$ and $\chi$ is then arbitrary, as in previous studies.
Another difference is that $\eta_I$ is not involved in the
one-loop neutrino mass, contrary to the original Majorana case of
Ref.~\cite{m06}. The above possibility has also been discussed in
\cite{gs08}. In the following, we introduce a new possibility for
dark matter candidate within the present scenario.

Since $m_{N_k}$ are assumed to be very large in this scenario, the
annihilation of the dark-matter scalars in this model do not
proceed via their Yukawa interactions, but rather through their
gauge or scalar interactions. Examples of the latter have been
discussed extensively in the
literature~\cite{sz85,m94,bptv01,pw06,blmrs08,aht08,aahlt10,fps10}.

We consider next the lightest $N_k$ (call it $N_1$) as dark
matter. As shown previously~\cite{kms06}, this is subject to
severe phenomenological constraints in the original
model~\cite{m06} of scotogenic Majorana neutrino mass.  The reason
is as follows.  In order for $N_1 N_1$ annihilation to account for
the correct relic abundance, the $\eta$ masses cannot be too heavy
and the Yukawa couplings $f_{\alpha k}$ cannot be too small.
However, these values are severely constrained by experimental
upper limits on the $\mu \to e \gamma$ rate, as already discussed.
It is thus not a viable option, without some detailed fine tuning
of parameters.  To retain $N_1$ as a natural dark-matter
candidate, new interactions involving $N_1$ need to be postulated,
such as a singlet scalar~\cite{bm08}.  In our present model, the
$h_{kj} N_k \nu^c_j \chi^0$ Yukawa couplings are exactly what are
required.  They are not constrained by flavor-changing
charged-lepton radiative decays, so they can be large enough for a
realistic $N_1 \bar{N}_1$ annihilation cross section to account
for the relic abundance of dark matter in the Universe today.  In
this scenario, the $f_{\alpha k}$ Yukawa couplings as well as the
$A$ parameter  are small and the mass of $\zeta_1$ (which is
mostly composed of $\chi$) is not much greater than $m_{N_1}$.

Combining $(\nu,\nu^c)$ and $(N,N^c)$ to form the four-component Dirac
fermions $\nu$ and $N$, their Yukawa interactions are given by
\begin{equation}
{\cal L}_Y = f_{\alpha k} \bar{N}_k \left( {1 - \gamma_5 \over 2} \right)
(\nu_\alpha \eta^0 - l_\alpha \eta^+) + h_{k \beta} \bar{N}_k \left(
{1 + \gamma_5 \over 2} \right) \nu_\beta \chi^0 + H.c.
\end{equation}
where in this four-component notation, $[(1+\gamma_5)/2]\nu$ represents
$\nu^c$ going backwards. For the dark-matter candidate $N_1$, we assume
$h_{1 \beta}$ to be dominant, then
\begin{equation}
\sigma(N_1+\bar{N}_1 \to \nu_\alpha +
\bar{\nu}_\beta)=\sum_{\alpha,\beta} \frac{|h_{1 \alpha}^* h_{1
\beta}|^2}{32\pi v_{rel}}\frac{m_{N_1}^2}
{(m_{N_1}^2+m_{\zeta_1}^2)^2} < \sum_{\alpha,\beta} \frac{|h_{1
\alpha}^* h_{1 \beta}|^2}{128\pi v_{rel} m_{N_1}^2},
\end{equation}
where to reach the last inequality we have used $ m_{\zeta_1}>
m_{N_1}$.  Similarly,
\begin{equation}
\sigma(N_1+{N}_1 \to \nu_\alpha + {\nu}_\beta)=\sum_{\alpha,\beta}
\frac{|h_{1 \alpha}^* h_{1 \beta}^*|^2}{32\pi
v_{rel}}\frac{m_{N_1}^2} {(m_{N_1}^2+m_{\zeta_1}^2)^2} <
\sum_{\alpha,\beta} \frac{|h_{1 \alpha}^* h_{1 \beta}^*|^2}{128\pi
v_{rel} m_{N_1}^2}
\end{equation}
Setting the sum of the two annihilation cross sections times the
relative velocity equal to one picobarn, we find
\begin{equation}
m_{N_1} < \left( \sum_{\alpha,\beta} |h_{1 \alpha}^* h_{1 \beta}|^2
\right)^{1/2} (1.4~{\rm TeV}).
\end{equation}
For $|h_{1 \alpha}| < 1$, we then obtain $m_{N_1} < 4.2$ TeV. With
such light $N_1$, the seesaw mechanism is not very effective.  The
smallness of the neutrino masses can be justified by the smallness
of the trilinear $A$ term which softly breaks $Z_2^{(A)}$, and the
smallness of the $f$  Yukawa couplings. If the $h$ couplings were
not available,  the cross section must have then come from the $f$
couplings, which are restricted by $\mu \to e \gamma$, so the
annihilation cross section would in general be too small
 for $N_1$ to be a viable dark-matter candidate.  If the
$B-L$ symmetry is gauged, there should be another annihilation
mode $ N + \bar{N} \to Z' \to \nu + \bar{\nu}, l + \bar{l}, q +
\bar{q}$. This cross section is given by~\cite{klm10}
\begin{equation}
\sigma = {g_{Z'}^4 m_{N_1}^2 \over \pi v_{rel} (4 m_{N_1}^2 - m_{Z'}^2)^2}
\end{equation}
The present lower bound on $m_{Z'}$ from the Large Hadron Collider
(LHC)~\cite{Zprime} is estimated to be about 2 TeV. For $g_{Z'} =
\sqrt{(5/8)} g_Y = 0.28$ (i.e. the $SO(10)$ limit), $m_{Z'} = 2$
TeV, and $\sigma v_{rel} = 1$ pb, we find $m_{N_1} = 900$ GeV.  In
this case, $N_1 \bar{N_1}$ production from $Z'$ decay at the LHC
is possible, as studied previously~\cite{hkor08}, except that
$N_1$ is now dark matter. It may however be inferred from the
increase of the $Z'$ invisible width on top of the expected $Z'
\to \nu \bar{\nu}$ mode. As $N_1$ is otherwise very difficult to
produce, the existence of $Z'$ seems to be the only realistic
chance for it to be observed at the LHC, but still only
indirectly. If $\eta^+$ is light enough, it can be produced at the
LHC.  The subsequent decay of $\eta^+$ into $N_1$ and a charged
lepton is a possible signature, as discussed in
Ref.~\cite{hashemi}.

As for direct detection of dark matter in underground experiments,
if $B-L$ is not gauged, then $N_1$ has no interaction with nuclei.
If $B-L$ is gauged, then the  elastic scattering of $N_1$ with
nuclei may proceed through $Z'$ exchange.  The cross section per
nucleon is given by~\cite{klm10}
\begin{equation}
\sigma_0 = {4 m_P^2 \over \pi} {g_{Z'}^4 \over m_{Z'}^4}.
\end{equation}
For $g_{Z'} = 0.28$ and $m_{Z'}=2$ TeV, this implies $\sigma_0 =
1.7 \times 10^{-7}$ pb, which exceeds the XENON100
bound~\cite{xenon11} of about $7 \times 10^{-8}$ pb for $m_{N_1} =
900$ GeV.  This means that in
this case, $g_{Z'}/m_{Z'}$ should be reduced by a factor of 1.25
or more.

This minimal model is also very suitable for the implementation of
the non-Abelian discrete $A_4$ symmetry~\cite{mr01} to the neutrino
mass matrix~\cite{m04}.  In the charged-lepton sector, let
$(\nu_i,l_i) \sim \underline{3}$ under $A_4$, and either $l^c_i
\sim \underline{1}, \underline{1}', \underline{1}''$ as in
Ref.~\cite{mr01} or $l^c_i \sim \underline{3}$ as in
Ref.~\cite{m06+1}, then with $\Phi \sim \underline{3}$ or
$\underline{3} + \underline{1}$, and $A_4$ breaking to the
residual symmetry $Z_3$, the charged-lepton mass matrix is
diagonalized by the well-known unitary matrix
\begin{equation}
U_L = {1 \over \sqrt{3}} \pmatrix{1 & 1 & 1 \cr 1 & \omega &
\omega^2 \cr 1 & \omega^2 & \omega},
\end{equation}
where $\omega = \exp(2 \pi i/3)$.  In the neutrino sector, let
$\nu^c_i \sim \underline{3}$, $\eta \sim \underline{1}$, and
$\chi \sim \underline{1} + \underline{3}$, with the soft scalar
trilinear $\chi \Phi^\dagger \eta$ terms to break $A_4$, the neutrino
mass matrix becomes~\cite{m04}
\begin{equation}
{\cal M}_\nu = \pmatrix{a & f & e \cr f & a & d \cr e & d & a}.
\end{equation}
If $e=f=0$, then neutrino mixing is tribimaximal, i.e. $\sin^2
\theta_{12} = 1/3$, $\sin^2 \theta_{23} = 1/2$, $\theta_{13}=0$.
This was known to be a good approximation of the measured neutrino
mixing angles.  However, two recent experiments have measured
$\theta_{13}$ to be definitely nonzero, i.e.
\begin{equation}
\sin^2 2 \theta_{13} = 0.092 \pm 0.016({\rm stat}) \pm 0.005({\rm syst})
\end{equation}
from the Daya Bay Collaboration~\cite{daya12}, and
\begin{equation}
\sin^2 2 \theta_{13} = 0.113 \pm 0.013({\rm stat}) \pm 0.019({\rm syst})
\end{equation}
from the RENO Collaboration~\cite{reno12}.
In that case, $e$ and $f$ should be nonzero.  Let
\begin{equation}
\epsilon = {e-f \over d \sqrt{2}}, ~~~~ \delta = {e+f \over d
\sqrt{2}}.
\end{equation}
The parameters $a,d,e,f$ are complex, and for small $e,f$, the
eigenvalues of ${\cal M}_\nu$ are $a+d$, $a$, and $a-d$.  We can
always choose $a$ to be real,  the phase of $d$ is then determined
by the absolute values of the three masses~\cite{m05}. For the
small values of $e$ and $f$, we find
\begin{equation}
\theta_{13} = -{\epsilon \over \sqrt{3}}, ~~~~~ \tan^2 \theta_{12}
= {1 \over 2} \left[ {(1 - \sqrt{2} Re\delta)^2 + 2 (Im\delta)^2 \over
(1 + Re\delta/\sqrt{2})^2 + (Im\delta)^2/2} \right].
\end{equation}
Thus, a nonzero $\theta_{13} $ and a value of $\tan^2 \theta_{12}$
smaller than $ 0.5$ can be obtained.  More precisely, the neutrino
mass matrix in the tribimaximal basis is now of the form
\begin{equation}
{\cal M}_\nu^{(1,2,3)} = \pmatrix{m_1 & m_6 & 0 \cr m_6 & m_2 &
m_5 \cr 0 & m_5 & m_3} = \pmatrix{a+d & \delta d & 0 \cr
\delta d & a & \epsilon d \cr 0 & \epsilon d &
a-d}.\label{m4zero}
\end{equation}
If $\delta=\epsilon=0$, the tribimaximal mixing is then
recovered. This differs from the originally proposed
deviation~\cite{m04} for $A_4$, which was updated
recently~\cite{mw11}, i.e.
\begin{equation}
{\cal M}_\nu^{(1,2,3)} = \pmatrix{m_1 & 0 & m_4 \cr 0 & m_2 & m_5
\cr m_4 & m_5 & m_3} = \pmatrix{a+d-(b+c)/2 & 0 & i\sqrt{3}/2(c-b)
\cr 0 & a+b+c & \sqrt{2}e \cr i\sqrt{3}/2(c-b) & \sqrt{2}e &
a-d-(b+c)/2}.
\end{equation}
Given that $m_4=0$ in Eq.~(\ref{m4zero}), we obtain the approximate
relationship
\begin{equation}
\sin^2 2 \theta_{23} \simeq 1 - 8 [Re(U_{e3})]^2.
\end{equation}
Using the experimental bound $\sin^2 2 \theta_{23} > 0.92$, we
find $|Re(U_{e3})| < 0.1$.  If we take the central value of
$|U_{e3}|$ to be 0.16 (corresponding to $\sin^2 2 \theta_{13} =
0.1$), we then obtain $|\tan \delta_{CP}| > 1.3$ in this model.
Details are given elsewhere~\cite{im12}.

Below we also mention briefly how a supersymmetric model of scotogenic
neutrino mass may be constructed.  Consider the superfield
content listed in Table 2.
\begin{table}[htb]
\begin{center}
\begin{tabular}{|c|c|c|c|c|c|c|}
\hline
superfields & $SU(3)_C$ & $SU(2)_L$ & $U(1)_Y$ & $U(1)_{B-L}$ &
$Z_2^{(A)}$
& $Z_2^{(B)}$ \\
\hline
$(u,d)$ & 3 & 2 & 1/6 & 1/3 & + & + \\
$u^c$ & $3^*$ & 1 & --2/3 & --1/3 & + & + \\
$d^c$ & $3^*$ & 1 & 1/3 & --1/3 & + & + \\
\hline
$(\nu,e)$ & 1 & 2 & --1/2 & --1 & + & + \\
$e^c$ & 1 & 1 & 1 & 1 & -- & + \\
$\nu^c$ & 1 & 1 & 0 & 1 & -- & + \\
\hline
$(\phi_1^0,\phi_1^-)$ & 1 & 2 & --1/2 & 0 & -- & + \\
$(\phi_2^+,\phi_2^0)$ & 1 & 2 & 1/2 & 0 & + & -- \\
$(\phi_3^0,\phi_3^-)$ & 1 & 2 & --1/2 & 0 & + & + \\
$(\phi_4^+,\phi_4^0)$ & 1 & 2 & 1/2 & 0 & + & + \\

\hline
$\chi_1^+$ & 1 & 1 & 1 & 0 & -- & + \\
$\chi_1^0$ & 1 & 1 & 0 & 0 & -- & + \\
$\chi_2^0$ & 1 & 1 & 0 & 0 & -- & -- \\
$\chi_2^-$ & 1 & 1 & --1 & 0 & + & -- \\
\hline
$N$ & 1 & 1 & 0 & --1 & + & -- \\
$N^c$ & 1 & 1 & 0 & 1 & + & -- \\
\hline
\end{tabular}
\caption{Assignments of the particles of the supersymmetric model
under $B-L$, $Z_2^{(A)}$ and $Z_2^{(B)}$.}
\end{center}
\end{table}
\begin{figure}[htb]
\begin{center}
\begin{picture}(480,100)(0,0)
\ArrowLine(60,10)(90,10) \ArrowLine(130,10)(90,10)
\ArrowLine(130,10)(170,10) \ArrowLine(200,10)(170,10)
\DashArrowArc(130,10)(40,90,180)5 \DashArrowArcn(130,10)(40,90,0)5
\DashLine(130,50)(130,80)5 \Text(75,0)[]{\large $\nu$}
\Text(185,0)[]{\large $\nu^c$} \Text(110,0)[]{\large $N^c$}
\Text(150,0)[]{\large $N$} \Text(100,52)[]{\large $\phi_2^0$}
\Text(160,52)[]{\large $\chi_2^0$} \Text(130,90)[]{\large $\langle
\phi_1^0 \rangle$} \Text(130,10)[]{\Large $\times$}
\ArrowLine(280,10)(310,10) \DashArrowLine(350,10)(310,10)5
\DashArrowLine(350,10)(390,10)5 \ArrowLine(420,10)(390,10)
\ArrowArc(350,10)(40,90,180) \ArrowArcn(350,10)(40,90,0)
\DashLine(350,50)(350,80)5 \Text(285,0)[]{\large $\nu$}
\Text(405,0)[]{\large $\nu^c$} \Text(330,0)[]{\large
$\tilde{N}^c$} \Text(370,0)[]{\large $\tilde{N}$}
\Text(320,52)[]{\large $\tilde{\phi}_2^0$} \Text(380,52)[]{\large
$\tilde{\chi}_2^0$} \Text(350,90)[]{\large $\langle \phi_1^0
\rangle$} \Text(350,10)[]{\Large $\times$}
\end{picture}
\end{center}
\caption{One-loop generation of Dirac neutrino mass in the
supersymmetric case.}
\end{figure}
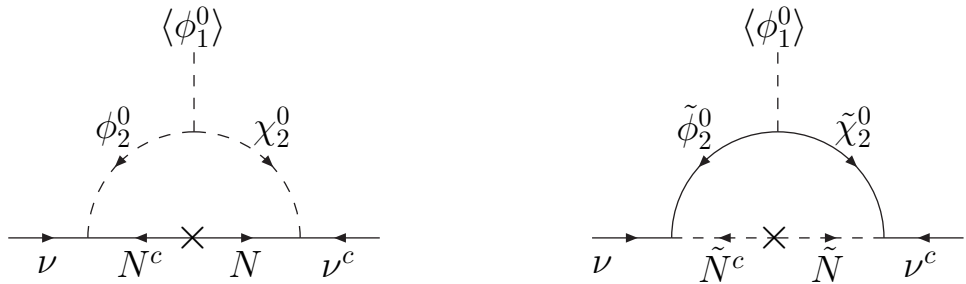
There are two one-loop diagrams contributing to the Dirac neutrino
mass as shown in Fig.~2.  Note that supersymmetry is broken by the
soft scalar trilinear $\chi_2^0 \phi_1^0 \phi_2^0$ and bilinear
$\tilde{N} \tilde{N}^c$ terms.  There are now many particles of odd
$Z_2^{(B)}$ as well as superpartners of odd $R$ parity.  There are
thus at least two dark-matter candidates~\cite{cmwy07}.  Obviously
the details of the dark sector are much more complicated.  We
will not study them further in this paper.

In conclusion, we have studied a minimal model of radiative Dirac
neutrino mass induced by dark matter.  In order for the scotogenic
Dirac neutrino mass to occur in one loop, we need to  introduce a
scalar singlet $\chi^0$ which mixes with the neutral component of
a new electroweak scalar doublet $(\eta^+,\eta^0)$. It is thus a
good theoretical justification for the existence of $\chi^0$. In
addition to the possibility of direct production at the LHC, the
presence of $\eta^+$ can modify the Higgs decay mode to $\gamma
\gamma$. As shown in \cite{carena}, if the $\lambda_3$ coupling in Eq.~(\ref{Vmain}) is negative, it can lead to the enhancement of Br($H \to \gamma \gamma$) in conformity of the recent observation at the LHC \cite{prepare}. Moreover,
the quartic coupling of $\chi^0$ with Higgs can stabilize its
potential against radiative corrections.

 This
minimal model also requires three heavy neutral Dirac fermions
$N_i$. Depending on the mass spectrum, the dark matter might be
either the lightest Dirac fermion $N_1$ or one of the neutral
scalars; i.e. the imaginary component $\eta_I$ of $\eta^0$
or a linear combination of the real component $\eta_R$ of $\eta^0$
and $\chi^0$. In the latter case, depending on the
mixing between $\eta_R$ and $\chi^0$, which should be small because of
the soft breaking of $Z_2^{(A)}$, the annihilation rate due to
the electroweak interactions can be made equal to about
1~pb which is a value dictated by the dark matter abundance in the
thermal dark matter scenario.

If $N_1$ is the dark-matter candidate, its annihilation can
proceed via its Yukawa coupling with the right-handed neutrinos
and $\chi^0$.  This is a possibility that does not exist within the scotogenic
Majorana neutrino mass model because in that case the bounds from  the $\mu \to e \gamma$
constraints restrict the annihilation cross section of the $N_1$ pair below the required value. At the LHC, $N_1$ can then be produced via the decay
of $\eta^+$ and $\eta^-$ along with a charged lepton
\cite{hashemi}.

 The
$B-L$ symmetry used to maintain the conservation of lepton number
can be gauged. In that case, the present LHC lower bound on
$m_{Z'}$ is about 2 TeV. The interaction with the $Z'$ boson
provides another route for the annihilation of the $N_1$ pair as
well as a portal for the interaction with quarks and hence direct
detection. The bound from the XENON100 experiment already
constrains the parameter space.

 This minimal model is also
suitable for implementing an $A_4$ symmetry in such a way that
nonzero $\theta_{13}$ and large $\delta_{CP}$ may be obtained.  We
have also  briefly mentioned how a supersymmetric extension can be
constructed.

The work of E.M. is supported in part by the U.~S.~Department of Energy
under Grant No.~DE-AC02-06CH11357.
The authors thank
Galileo Galilei Institute for Theoretical Physics for its hospitality.
 YF acknowledges  partial support from
the  European Union FP7  ITN INVISIBLES (Marie Curie Actions,
PITN- GA-2011- 289442). She is also grateful to ICTP for partial
financial support and hospitality.

\bibliographystyle{unsrt}

\end{document}